\documentclass[12pt]{article}

\usepackage{latexsym}

\usepackage{graphicx}

\begin{document}


\title{Thermodynamics of rotating charged dilaton black holes  in an external magnetic field}

\author{
     Stoytcho S. Yazadjiev \thanks{E-mail: yazad@phys.uni-sofia.bg}\\
{\footnotesize  Department of Theoretical Physics,
                Faculty of Physics, Sofia University,}\\
{\footnotesize  5 James Bourchier Boulevard, Sofia~1164, Bulgaria
}\\  }

\date{}

\maketitle

\begin{abstract}
In the present paper we study the long-standing problem for the
thermodynamics of magnetized  dilaton black holes. For this purpose
we construct an exact solution describing a rotating charged dilaton
black hole immersed in an external magnetic field and discuss its
basic properties. We derive a Smarr-like relation and the
thermodynamics first law for these magnetized black holes. The
novelty in the thermodynamics of the magnetized black holes is the
appearance of new terms proportional to the magnetic momentum of the
black holes in the Smarr-like relation and the first law.

\end{abstract}


\sloppy

\section{Introduction}

Black holes influenced by external fields are interesting subject to
study and have stimulated  a lot of research during the last three
decades. Black holes immersed in external magnetic field (the
so-called magnetized black holes) are of particular interest to
astrophysics where the magnetic field plays important role in the
physical processes taking place in the black hole vicinity
\cite{WALD}--\cite{Y4}. The study of black holes in external
magnetic field results in finding interesting astrophysical effects
as the charge accretion and the flux expulsion from extreme black
holes \cite{WALD}, \cite{EW}, \cite{KARAS}--\cite{KV1}, \cite{KB},
\cite{TNIKY}. From a theoretical point of view, it is very
interesting to investigate how the external fields influence the
black hole thermodynamics -- more specifically, how the external
magnetic field affects it.

The  expectation was that a thermodynamic description of the black
holes in external magnetic field would include also the value of the
background magnetic field as a further parameter. However, more
detailed studies showed that this was not the case for  static
electrically uncharged solutions in four dimensions and some
electrically uncharged rotating black holes in higher dimensions --
the external magnetic field only distorts the horizon geometry
without affecting the thermodynamics \cite{RADU}, \cite{O1},
\cite{Yazadjiev1}. In a recent paper \cite{Y4} the author studied
the thermodynamics of some charged dilaton black holes in external
magnetic  field and he found that even in this case the black hole
thermodynamics is not affected by the external magnetic field.

A natural and a long-standing problem is whether this remains true
in the rotating case. More or less surprisingly this problem has not
been investigated systematically in the literature even for the
magnetized Kerr-Newman solution in Einstein-Maxwell gravity. One
possible reason for this is the fact that the magnetized Kerr-Newman
solution is rather complicated.

The main purpose of the present paper is to study the long-standing
problem  for the thermodynamics of the magnetized rotating and
electrically charged black holes. More precisely we study the
thermodynamics of electrically charged and rotating black holes in
Einstein-Maxwell-dilaton (EMd) gravity. For this purpose we
construct an exact solution describing a rotating charged dilaton
black hole with a dilaton coupling parameter $\alpha=\sqrt{3}$
immersed in an external magnetic field and discuss its basic
properties. Then we show how the external magnetic field affects the
thermodynamics by deriving  a Smarr-like relation and the first law.
The novelty in the thermodynamics of the magnetized black holes is
the appearance of new terms proportional to the magnetic momentum of
the black holes in the Smarr-like relation and the first law.

\section{Exact rotating  charged dilaton black hole in  an external magnetic field}

The  EMd gravity is described by the following field equations

\begin{eqnarray}
&&R_{\mu\nu}= 2\nabla_{\mu}\varphi \nabla_{\nu}\varphi  + 2e^{-2\alpha\varphi}  \left(F_{\mu\sigma}F_{\nu}{}^{\sigma}- \frac{g_{\mu\nu}}{4}F_{\rho\sigma}F^{\rho\sigma} \right),\\
&&\nabla_{\mu}\left(e^{-2\alpha\varphi}F^{\mu\nu}\right)=0=\nabla_{[\mu} F_{\nu\sigma]}, \\
&&\nabla_{\mu}\nabla^{\mu}\varphi=
-\frac{\alpha}{2}e^{-2\alpha\varphi}F_{\rho\sigma}F^{\rho\sigma} ,
\end{eqnarray}
where  $\nabla_{\mu}$ and  $R_{\mu\nu}$ are the  Levi-Civita
connection and the Ricci tensor with respect to the spacetime metric
$g_{\mu\nu}$. $F_{\mu\nu}$ is the Maxwell tensor and the dilaton
field is denoted by $\varphi$, with $\alpha$ being the dilaton
coupling parameter governing the coupling strength of the dilaton to
the electromagnetic field.

The asymptotically flat, rotating and  electrically charged dilaton
black holes with $\alpha=\sqrt{3}$ are described by the following
exact solution to the EMd equations \cite{FZ}

\begin{eqnarray}
&&ds^2_{0}= \left[H(r^2 + a^2) + a^2H^{-1}Z\sin^2\theta \right]
\sin^2\theta \left(d\phi - \frac{a}{\sqrt{1-\upsilon^2}}
\frac{H^{-1}Z}{\left[H(r^2 + a^2) + a^2H^{-1}Z\sin^2\theta \right]}
dt\right)^2 \nonumber  \\
&&- \left(H^{-1}(1-Z) + \frac{a^2}{1-\upsilon^2} \frac{H^{-2}Z^2
\sin^2\theta }{\left[H(r^2 + a^2) + a^2H^{-1}Z\sin^2\theta \right]}
\right)dt^2 + H \frac{\Sigma}{\Delta} dr^2 + H\Sigma d\theta^2,
\end{eqnarray}

\begin{eqnarray}
A^{0}_{t}= \frac{\upsilon}{2(1-\upsilon^2)}H^{-2}Z,
\end{eqnarray}

\begin{eqnarray}
A^{0}_{\phi}= -\frac{a\upsilon}{2\sqrt{1-\upsilon^2}} H^{-2}Z
\sin^2\theta ,
\end{eqnarray}

\begin{eqnarray}
e^{\frac{2}{3}\sqrt{3}\varphi_{0}}=H^{-1},
\end{eqnarray}
where the functions $H$, $Z$, $\Sigma$ and $\Delta$ are given by

\begin{eqnarray}
H=\sqrt{\frac{1- \upsilon^2 + \upsilon^2 Z}{1-\upsilon^2}},   \;\;\;
Z=  \frac{2mr}{\Sigma},
\end{eqnarray}

\begin{eqnarray}
\Sigma= r^2 + a^2\cos^2\theta, \;\;\; \Delta= r^2 - 2mr + a^2.
\end{eqnarray}
The solution parameters $\upsilon$, $m$ and $a$ are related to the
mass $M_{0}$, the angular momentum $J_{0}$ and the charge $Q_{0}$ of
the black hole via the formulae

\begin{eqnarray}
M_{0}=m\left(1 + \frac{1}{2}\frac{\upsilon^2}{1- \upsilon^2}\right),
\;\; J_{0}=\frac{ma}{\sqrt{1- \upsilon^2}}, \;\; Q_{0}=
\frac{m\upsilon}{1-\upsilon^2} .
\end{eqnarray}

The event horizon lies at
\begin{eqnarray}
r_{+}= m + \sqrt{m^2 - a^2 },
\end{eqnarray}
which is the greater root of $\Delta=0$ and the  angular velocity of
the horizon is

\begin{eqnarray}
\Omega^{\,0}_{H}= \sqrt{1- \upsilon^2} \frac{a}{r^2_{+} + a^2} .
\end{eqnarray}

Another very important characteristic of the black hole, which will
play a crucial role in the thermodynamics of the magnetized black
holes, is the magnetic momentum $\mu$ defined from the asymptotic
behavior of $A^{0}_{\phi}$ for $r\to \infty$, namely
\begin{eqnarray}\label{mm1}
A^{0}_{\phi}\to - \mu \frac{\sin^2\theta}{r}
\end{eqnarray}
and  given by
\begin{eqnarray}\label{mm2}
\mu=\frac{ma\upsilon}{\sqrt{1-\upsilon^2}}= \upsilon
J_{0}=\sqrt{1-\upsilon^2}a Q_{0} .
\end{eqnarray}

In order to construct the exact solution describing rotating and
charged dilaton black holes in an external magnetic field we shall
follow the following scheme. We consider the 4D rotating charged
dilaton black holes with a metric
\begin{eqnarray}
ds^2_{0}=g^{0}_{\mu\nu}dx^\mu dx^\nu=X_{0}(d\phi + W^{0} dt)^2 +
X^{-1}_{0}\gamma_{ab}dx^a dx^b,
\end{eqnarray}
scalar field $\varphi_{0}$ and gauge potential $A^{0}_{\mu}$. Its
Kaluza-Klein uplifting to a 5D vacuum Einstein solution is given by

\begin{eqnarray}
ds^2_{5}=e^{\frac{2}{3}\sqrt{3}\varphi_{0}} ds^2_{0} +
e^{-\frac{4}{3}\sqrt{3}\varphi_{0}}\left(dx_{5} +
2A^{0}_{\mu}dx^\mu\right)^2 .
\end{eqnarray}
Then we perform a twisted Kaluza-Klein reduction along the Killing
field $V=B\frac{\partial}{\partial \phi} + \frac{\partial}{\partial
x_{5}}$, which gives the following 4D solution to the EMd equations
with $\alpha=\sqrt{3}$:
\begin{eqnarray}
ds^2 = \Lambda^{-1/2} X_{0}(d\phi + W dt)^2 + \Lambda^{1/2}
X^{-1}_{0}\gamma_{ab}dx^a dx^b,
\end{eqnarray}

\begin{eqnarray}
W= (1+ 2BA^{0}_{\phi})W^{0} - 2B A^{0}_{t},
\end{eqnarray}

\begin{eqnarray}
A_{\phi}= \Lambda^{-1}(1+ 2BA^{0}_{\phi})A^{0}_{\phi} +
\frac{B}{2}\Lambda^{-1} X_{0}e^{2\sqrt{3}\varphi_{0}},
\end{eqnarray}

\begin{eqnarray}
A_{t}= \Lambda^{-1}(1+ 2BA^{0}_{\phi})A^{0}_{t} +
\frac{B}{2}\Lambda^{-1} X_{0}e^{2\sqrt{3}\varphi_{0}} W^{0},
\end{eqnarray}

\begin{eqnarray}
e^{\frac{4}{3}\sqrt{3}\varphi}= \Lambda^{-1}
e^{\frac{4}{3}\sqrt{3}\varphi_{0}},
\end{eqnarray}
where $\Lambda$ is given by
\begin{eqnarray}
\Lambda = \left(1 + 2BA^{0}_{\phi}\right)^2 + B^2
X_{0}e^{2\sqrt{3}\varphi_{0}} .
\end{eqnarray}
Alternative derivation of the rotating and electrically charged
dilaton black hole solution is given in the appendix.

We start investigating the basic properties of the  magnetized
solution by noting that there is no conical singularity on the axis
of symmetry since both $A^{0}_{\phi}$ and $X_{0}$ vanish at the
axis. This means that the periodicity of the angular coordinate
$\phi$ is the standard one $\Delta\phi=2\pi$. As in the non-rotating
case the parameter $B$ is the asymptotic strength of the magnetic
field. The inspection of the metric shows that there is a Killing
horizon at $r=m + \sqrt{m^2-a^2}$ where the Killing field
$K=\frac{\partial}{\partial t} + \Omega_{H}\frac{\partial}{\partial
\phi}$ becomes null. Here $\Omega_{H}$ is the angular velocity of
the horizon given by

\begin{eqnarray}\label{OH}
\Omega_{H}= - W|_{H}= - W^{0}|_{H} + 2B (A^{0}_{t}-
A^{0}_{\phi}W^{0})_{H}= \nonumber \\  \Omega^{0}_{H}  + 2B(A^{0}_{t}
+ \Omega^{0}_{H}A^{0}_{\phi})|_{H} =\Omega^{0}_{H} +  2B
\Xi^{0}_{H}= \Omega^{0}_{H} + B\upsilon ,
\end{eqnarray}
where $\Omega^{0}_{H}$ is the horizon angular velocity and
$\Xi^{0}_{H}=(A^{0}_{t} +
\Omega^{0}_{H}A^{0}_{\phi})|_{H}=\frac{1}{2}\upsilon$ is the
corotating electric potential evaluated on the horizon for the
non-magnetized seed solution.

The metric induced on the horizon cross section is
\begin{eqnarray}\label{HM}
&&ds^2_{H}= \Lambda^{-1/2}X_{0}d\phi^2 + \Lambda^{1/2}
g^{(0)}_{\theta\theta}d\theta^2 =  \nonumber\\ &&
\Lambda^{-1/2}\left[H(r^2 + a^2) + a^2H^{-1}Z\sin^2\theta \right]
\sin^2\theta d\phi^2 + \Lambda^{1/2}H\Sigma d\theta^2 .
\end{eqnarray}
The horizon area calculated via (\ref{HM}) coincides with the
horizon area of the non-magnetized seed solution and is given by
\begin{eqnarray}
{\cal A}_{H}={\cal A}^{0}_{H}=\frac{4\pi(r^2_{+} +
a^2)}{\sqrt{1-\upsilon^2}} .
\end{eqnarray}

Although the horizon cross section is a topological 2-sphere, it is
geometrically deformed by the rotation and  the external magnetic
field. The horizon surface gravity $\kappa$ can be found by the
well-known formula
\begin{eqnarray}
\kappa^2 = -\frac{1}{4\lambda} g^{\mu\nu}\partial_{\mu}\lambda
\partial_{\nu}\lambda ,
\end{eqnarray}
where $\lambda=g(K,K)$. By direct computation we find that the
surface gravity of the magnetized solution coincides with the
surface gravity of the seed solution

\begin{eqnarray}
\kappa=\kappa_{0}= \sqrt{1- \upsilon^2}\,
\frac{r_{+}-r_{-}}{(r^2_{+} + a^2)} =\sqrt{1- \upsilon^2}\,
\frac{\sqrt{m^2-a^2}}{(r^2_{+} + a^2)} .
\end{eqnarray}

The solution possesses an ergoregion determined by the inequality
$g(\frac{\partial}{\partial t},\frac{\partial}{\partial
t})=g_{tt}>0$. The explicit form of $g_{tt}$ can be presented as
follows

\begin{eqnarray}
g_{tt}= \sqrt{\Lambda} \left[g^{0}_{tt} - X_{0}(W^{0})^2 +
\frac{X_{0}}{\Lambda}W^2 \right].
\end{eqnarray}
Using this form of $g_{tt}$, it is not difficult to see that very
close to the horizon we have $g_{tt}\ge0$ and $g_{tt}<0$ for large
$r$. Therefore, we conclude that there is an ergoregion confined in
a compact neighborhood of the horizon, in contrast with the
magnetized Kerr-Newman solution for which the ergoregion extends to
infinity \cite{GMP}.

Concerning the asymptotic of the magnetized solution, it is not
difficult to see from  its explicit form that the solution
asymptotes the static dilaton-Melvin solution with
$\alpha=\sqrt{3}$. This is also confirmed by inspecting the
asymptotic behavior of the electric field $\vec E$ and the twists
$\omega_{\eta}=i_{\eta}\star d\eta$ and $\omega_{\xi}=i_{\xi}\star
d\xi$ associated with the Killing vectors
$\eta=\frac{\partial}{\partial \phi}$ and $\xi=
\frac{\partial}{\partial t}$. For these quantities we have ${\vec
E}^2 \to 0$, $\omega_{\eta} \to 0$ and $\omega_{\xi}\to 0$ for $r\to
\infty$.

\section{Thermodynamics}

The main purpose of the present paper is to study the thermodynamics
of the magnetized rotating, charged dilaton black holes. The first
step in this direction is to find the conserved physical quantities,
namely the electric charge, the mass and the angular momentum
associated with the magnetized black hole. The physical electric
charge can be calculated as usual via the well-known formula and the
result is

\begin{eqnarray}\label{CHARGE}
Q=\frac{1}{4\pi} \int_{H} e^{-2\sqrt{3}\varphi}\star F = Q_{0} -
2BJ_{0}.
\end{eqnarray}

The calculation of the mass and the angular momentum, however, is
much more tricky since the spacetime is not asymptotically flat -- a
substraction procedure is needed to obtain finite quantities from
integrals divergent at infinity. The natural choice for the
substraction background in our case is the static dilaton-Melvin
background. To calculate the mass we use the quasilocal formalism
[23]. Here we give for completeness a very brief description of the
quasilocal formalism. We foliate the spacetime by spacelike surfaces
$\Sigma_{t}$ of metric $h_{\mu\nu} = g_{\mu\nu} + u_{\mu}u_{\nu}$,
labeled by a time coordinate $t$ with a unit normal vector $u^{\mu}
= - N\delta^{\mu}_{0}$. The spacetime metric is then decomposed into
the form

\begin{equation}
ds^2 = - N^2dt^2 + h_{ij}(dx^i + N^{i}dt)(dx^j + N^{j}dt),
\end{equation}
with $N$ and $N^{i}$ being the lapse function and  the shift vector.

The spacetime boundary consists of the initial surface $\Sigma_{i}$
($t=t_{i}$), the final surface $\Sigma_{f}$ ($t=t_{f}$) and a
timelike surface ${\cal B}$ to which the vector $u^{\mu}$ is
tangent. The surface ${\cal B}$ is foliated by $2$-dimensional
surfaces $S^{r}_{t}$, with metric $\sigma_{\mu\nu}= h_{\mu\nu} -
n_{\mu}n_{\nu}$, which are the intersections of $\Sigma_{t}$ and
${\cal B}$. The unit spacelike outward normal to $S^{r}_{t}$,
$n_{\mu}$, is orthogonal to $u^{\mu}$.

Let us denote by $K$  the trace of the extrinsic curvature
$K^{\mu\nu}$ of $\Sigma_{t_{i,f}}$ and by $\Theta$  the trace of the
extrinsic curvature $\Theta^{\mu\nu}$ of ${\cal B}$, given by

\begin{eqnarray}
K_{\mu\nu} &=& - {1\over 2N}\left( {\partial h_{\mu\nu}\over \partial t} - 2D_{(\mu}N_{\nu)} \right) ,\\
\Theta_{\mu\nu} &=& - h^{\alpha}_{\mu} \nabla_{\alpha} n_{\nu},
\end{eqnarray}
where $\nabla_{\mu}$ and $D_{\nu}$ are the covariant derivatives
with respect to the metric $g_{\mu\nu}$ and $h_{\mu\nu}$,
respectively. The quasilocal energy $M$ and the angular momentum
$J_{i}$ are given by

\begin{eqnarray}
M = {1\over 8\pi} \int_{S^{r}_{t}} \sqrt{\sigma} \left[N(k-k_{*}) +
{n_{\mu}p^{\mu\nu}N_{\nu}\over \sqrt{h}} \right] d^{D-2}x \nonumber
\\ +
{1\over 4\pi}\int_{S^{r}_{t}} A_{t} \left({\hat \Pi}^{j} - {\hat \Pi}_{*}^{j} \right)n_{j}d^{D-2}x ,\\
J_{i} = - {1\over 8\pi} \int_{S^{r}_{t}} {n_{\mu}p^{\mu}_{i}\over
\sqrt{h} } \sqrt{\sigma}d^{D-2}x - {1\over 4\pi} \int_{S^{r}_{t}}
A_{i}{\hat \Pi}^{j}n_{j}d^{D-2}x .
\end{eqnarray}

Here $k= - \sigma^{\mu\nu}D_{\nu}n_{\mu}$ is the trace of the
extrinsic curvature of $S^{r}_{t}$ embedded in $\Sigma_{t}$. The
momentum variable $p^{ij}$ conjugated to $h_{ij}$ is given by

\begin{equation}
p^{ij} = \sqrt{h}\left(h^{ij}K - K^{ij} \right).
\end{equation}

The quantity ${\hat \Pi}^{j}$ is defined by

\begin{equation}
{\hat \Pi}^{j} = - {\sqrt{\sigma}\over \sqrt{h}}
\sqrt{-g}e^{-2\alpha\varphi}F^{tj}.
\end{equation}

The quantities with the subscript ``${}_{*}$'' are those associated
with the background.

Using the above formulae for the quasilocal mass and the quasilocal
angular momentum, after rather long but straightforward calculations
we find the following result for the total mass and the total
angular momentum

\begin{eqnarray}
M=M_{0}, \;\;\; J=J_{0}.
\end{eqnarray}
This result, combined with (\ref{OH}) and (\ref{CHARGE}), gives the
following Smarr-like relation for the magnetized black hole

\begin{eqnarray}\label{Smarr}
M= \frac{1}{4\pi}\kappa {\cal A}_{H} + 2\Omega_{H}J + \Xi_{H}Q - \mu
B ,
\end{eqnarray}
where  $\mu=2\Xi^{\,0}_{H}J_{0}$ is the magnetic moment of the
non-magnetized black hole defined by (\ref{mm1}) and (\ref{mm2}),
and $\Xi_{H}= K^{\mu}A_{\mu}|_{H}$ is the corotating electric
potential of the magnetized black hole evaluated on the horizon. It
is not difficult one to show that $\Xi_{H}=\Xi^{\,0}_{H}$, which
gives $\mu=2\Xi^{\,0}_{H}J_{0}=2\Xi_{H}J$.

For the first law of the magnetized black holes we obtain the
following expression

\begin{eqnarray}\label{FLMBH}
\delta M = \frac{\kappa}{2\pi} \delta \left(\frac{{\cal
A}_{H}}{4}\right) + \Omega_{H}\delta J + \Xi_{H}\delta Q
+{\mu}\delta B .
\end{eqnarray}

The first law can be easily checked by taking into account the fact
that  $\kappa=\kappa_{0}$ and ${\cal A}_{H}={\cal A}^{0}_{H}$, the
explicit expressions for $Q$ and $\Omega_{H}$, as well as the first
law for the non-magnetized black holes $\delta M_{0}=
\frac{\kappa_{0}}{2\pi}\delta \left(\frac{{\cal
A}^{0}_{H}}{4}\right) + \Omega^{0}_{H}\delta J_{0} +
\Xi^{0}_{H}\delta Q_{0}$.

From the explicit form of the Smarr-like relation  (\ref{Smarr}) and
the first law (\ref{FLMBH}) we see that the thermodynamics  of the
magnetized rotating electrically charged dilaton black holes indeed
includes the background magnetic field as a further parameter
conjugated to the magnetic moment of the black holes.  The way in
which the external magnetic field enters the Smarr-like relation and
the first law is what should be expected from a physical point of
view. The additional terms associated with the background magnetic
field in the Smarr-like relation and the first law  are in fact
related to the energy of the black hole considered as a magnetic
dipole in external magnetic field. This also explains the fact why
the static, neutral and electrically charged, magnetized black holes
have thermodynamic unaffected by the external field. In order for
the background magnetic field to affect the black hole
thermodynamics, the magnetic moment has to be non-zero  and this
occurs only for black holes which are simultaneously electrically
charged and with non-zero angular momentum according to the formula
$\mu=2\Xi^{0}_{H}J_{0}=2\Xi_{H}J$.

\section{Discussion}

In the present paper we derived the Smarr-like relation and the
first law for the magnetized rotating and electrically charged
dilaton black holes with dilaton coupling parameter
$\alpha=\sqrt{3}$. Taking into account the discussion at the end of
the previous section, on a physical background one should expect
that the Smarr-like relation and the first law for magnetized
rotating and electrically charged black holes with an arbitrary
dilaton coupling parameter $\alpha$ (including the Kerr-Newman
solution corresponding to $\alpha=0$) would have the same form as in
the case $\alpha=\sqrt{3}$. The systematic derivation, however,
seems to be nontrivial because in the general case the magnetized
solutions would not asymptote to the static dilaton-Melvin solution,
as it was demonstrated for the magnetized Kerr-Newman solution in
\cite{HIS} and \cite{GMP}. In other words, in order to obtain
meaningful expressions for  the mass and the angular momentum  of
the magnetized black hole solutions,  the substraction procedure has
to be carefully chosen in order to cope with the complicated
asymptotic behavior of the solutions. In fact there are some results
about the first law for the magnetized Kerr-Newman solution given in
\cite{Esteban}. However, the mass and angular momentum for the
magnetized Kerr-Newman black hole  were chosen ad hock in
\cite{Esteban} without a systematic derivation. We intend to
discussed these problems and their solution in a future publication.

\vspace{1.5ex}
\begin{flushleft}
\large\bf Acknowledgments
\end{flushleft}
The author is grateful to the Research Group Linkage Programme of
the Alexander von Humboldt Foundation for the support of this
research and the Institut f\"ur Theoretische Astrophysik T\"ubingen
for its kind hospitality. He also acknowledges partial support from
the Bulgarian National Science Fund under Grant DMU-03/6.

\vskip 1cm

\appendix

\section{Alternative derivation of the magnetized \\rotating charged dilaton black hole solution with $\alpha=\sqrt{3}$ }

We  consider stationary and axisymmetric EMd spacetimes, i.e
spacetimes  admitting a Killing field $\xi$  which is timelike at
infinity,  and a spacelike axial Killing field $\eta$ with closed
orbits. The dimensional reduction of the field equations along the
axial Killing field $\eta$ was performed in \cite{Y2}. So we present
here only the basic steps skipping the details which can be found in
\cite{Y2}. The invariance  of the Maxwell 2-form $F$  under  the
flow of the Killing field $\eta$ implies the existence of the
potentials  $\Phi$ and $\Psi$ defined by $d\Phi=i_{\eta}F$ and
$d\Psi=-e^{-2\alpha\varphi} i_{\eta}\star F$ and  such that

\begin{eqnarray}\label{Maxwell}
F=X^{-1}\eta \wedge d\Phi - X^{-1} e^{2\alpha\varphi} \star
\left(d\Psi \wedge \eta\right),
\end{eqnarray}
where

\begin{eqnarray}
X=g(\eta,\eta).
\end{eqnarray}
The twist $\omega_{\eta}$ of the Killing field $\eta$, defined by
$\omega_{\eta}=\star(d\eta\wedge \eta)$,  satisfies the equation

\begin{eqnarray}
d\omega_{\eta}= 4d\Psi \wedge d\Phi= d(2\Psi d\Phi - 2\Phi d\Psi).
\end{eqnarray}
Hence we conclude that there exists a twist potential $\chi$ such
that $\omega_{\eta}=d\chi + 2\Psi d\Phi - 2\Phi d\Psi$.

The projection metric $\gamma$ orthogonal to the Killing field
$\eta$, is defined by

\begin{eqnarray}
g= X^{-1}\left(\gamma + \eta\otimes \eta\right).
\end{eqnarray}
In local coordinates adapted to the Killing field, i.e.
$\eta=\partial/\partial\phi$, we have

\begin{eqnarray}
ds^2= g_{\mu\nu}dx^{\mu}dx^{\nu}= X(d\phi + W_a dx^a)^2 +
X^{-1}\gamma_{ab}dx^a dx^b.
\end{eqnarray}
The 1-form ${\cal W}=W_a dx^a$ is closely related to the twist
$\omega$ which can be expressed in the form

\begin{eqnarray}\label{TRF}
\omega_{\eta}= X i_{\eta} \star d{\cal W} .
\end{eqnarray}

The dimensionally reduced EMd equations for $\alpha=\sqrt{3}$ form
an effective 3-dimensional gravity coupled to a nonlinear matrix
$SL(3, R)/O(3)$ $\sigma$-model with the following action

\begin{eqnarray}
{\cal A}= \int d^{\,3}x \sqrt{-\gamma} \left[R(\gamma) -
\frac{1}{4}\gamma^{ab}  Tr\left(S^{-1}\partial_{a}S
S^{-1}\partial_{b}S\right) \right],
\end{eqnarray}
where $R(\gamma)$ is the Ricci scalar curvature with respect to the
metric $\gamma_{ab}$ and  $S$ is  a symmetric $SL(3,R)$ matrix
explicitly given by \cite{Y3}

\begin{eqnarray}
S\!=\!e^{-\frac{2\sqrt{3}}{3}\varphi}\!\!\left(\!\!
  \begin{array}{ccc}
  \!\! X^{-1}& -2X^{-1}\Phi & X^{-1}(2\Phi\Psi - \chi) \\
   \!\!  -2X^{-1}\Phi  &  e^{2\sqrt{3} \varphi} + 4X^{-1}\Phi^2 & -2e^{2\sqrt{3}\varphi}\Psi - 2X^{-1}(2\Phi\Psi -\chi)\Phi \\
   \!\! X^{-1}(2\Phi\Psi - \chi) & -2e^{2\sqrt{3}\varphi}\Psi - 2X^{-1}(2\Phi\Psi - \chi)\Phi &  X + 4\Psi^2e^{2\sqrt{3}\varphi} + X^{-1}(2\Phi\Psi -\chi)^2 \\
  \end{array}
\!\right) .\nonumber
\end{eqnarray}

The group of symmetries $SL(3,R)$ can be used to generate new
solutions from known ones via the scheme

\begin{eqnarray}
S \to \Gamma S \Gamma^{\,T},  \;\; \gamma_{ab} \to \gamma_{ab},
\end{eqnarray}
where $\Gamma\in SL(3,R)$. In the present paper we consider seed
solutions with potentials $(X_{0}, \Phi_{0}, \Psi_{0},
\chi_{0},\varphi_{0})$ corresponding to the seed matrix $S_{0}$ and
transformation matrices in the form

\begin{eqnarray}
\Gamma= \left(
            \begin{array}{ccc}
              1 & B & 0 \\
              0 & 1 & 0 \\
              0 & 0 & 1 \\
            \end{array}
          \right),
\end{eqnarray}
with $B$ being an arbitrary real number.

New solutions which can be generated from the seed  and the
transformation matrices under consideration, are encoded in the
matrix $S=\Gamma S_{0}\Gamma^{\,T}$ and the explicit form of their
potentials is given by

\begin{eqnarray}
&&X=\frac{X_{0}}{\sqrt{\Lambda}}, \\
&&\Phi= \Lambda^{-1}\Phi_{0}(1-2B\Phi_{0})  -\frac{B}{2} \Lambda^{-1} e^{2\sqrt{3}\varphi_{0}}X_{0},  \\
&&\Psi=(1- B\Phi_{0})\Psi_{0} + \frac{B}{2}\chi_{0},\\
&&\chi= \Lambda^{-1}\left\{(1-2B\Phi_{0})\chi_{0} + Be^{2\sqrt{3}\varphi_{0}}X_{0} \Psi_{0} \right. \nonumber \\
&&\left. + \left[B\Phi_{0}(1-2B\Phi_{0}) - \frac{1}{2}B^2e^{2\sqrt{3}\varphi_{0}}X_{0} \right](\chi_{0}- 2\Phi_{0}\Psi_{0}) \right\}, \\
&&e^{\frac{4}{\sqrt{3}}\varphi}= \Lambda^{-1}
e^{\frac{4}{\sqrt{3}}\varphi_{0}} ,
\end{eqnarray}
where $\Lambda=(1-2B\Phi_{0})^2 + B^2X_{0}e^{2\sqrt{3}\varphi_{0}}$.

In order to generate the solution describing the magnetized rotating
and electrically charged dilaton black holes we have to choose the
seed solution to be the asymptotically flat rotating electrically
charged dilaton black hole solution with $\alpha=\sqrt{3}$ found in
\cite{FZ}.

\end{document}